\documentclass[twocolumn,prl,superscriptaddress,showpacs]{revtex4}

\usepackage[T1]{fontenc}
\usepackage{times} % 11 pt, see \begin{document}

\usepackage{amsmath,amssymb,graphicx}
\usepackage{color,textcomp}

\renewcommand\epsilon{\varepsilon}
\renewcommand\phi{\varphi}
\renewcommand\theta{\vartheta}
\renewcommand\vec[1]{\boldsymbol{\mathrm{#1}}}
\newcommand\unitvec[1]{\vec{\hat {#1}}}
\newcommand\expect[1]{\left\langle\vphantom{\big(}#1\right\rangle}

\newcommand\Drot{D_\text{rot}}
\newcommand\tcoll{\tau_\text{coll}}
\newcommand\td{\tau_\text{d}}
\newcommand\trot{\tau_\text{rot}}
\DeclareMathOperator\pE{pE}

\newcommand\fig[1]{Fig.~\ref{fig:#1}}

\begin{document}
% adjust font size: 55 characters per 3.4in line, 57 lines per page
\makeatletter
\def\normalsize{%
    \@setfontsize\normalsize\@xipt{12}%
    \abovedisplayskip 10\p@ \@plus2\p@ \@minus5\p@
    \belowdisplayskip \abovedisplayskip
    \abovedisplayshortskip  \abovedisplayskip
    \belowdisplayshortskip \abovedisplayskip
    \let\@listi\@listI
}%
\makeatother
\normalsize

\title{Entangled Dynamics of a Stiff Polymer}

\author{Felix H{\"o}f\/ling}
\affiliation{Arnold Sommerfeld Center for Theoretical Physics (ASC)  and Center for
NanoScience (CeNS), Fakult{\"a}t f{\"u}r Physik,
Ludwig-Maximilians-Universit{\"a}t M{\"u}nchen, Theresienstra{\ss}e 37,
80333 M{\"u}nchen, Germany}
\affiliation{Hahn-Meitner-Institut,
Abteilung Theorie, Glienicker Stra{\ss}e 100, 14109 Berlin, Germany}
\author{Tobias Munk}
\affiliation{Arnold Sommerfeld Center for Theoretical Physics (ASC)  and Center for
NanoScience (CeNS), Fakult{\"a}t f{\"u}r Physik,
Ludwig-Maximilians-Universit{\"a}t M{\"u}nchen, Theresienstra{\ss}e 37,
80333 M{\"u}nchen, Germany}
\author{Erwin Frey}
\affiliation{Arnold Sommerfeld Center for Theoretical Physics (ASC)  and Center for
NanoScience (CeNS), Fakult{\"a}t f{\"u}r Physik,
Ludwig-Maximilians-Universit{\"a}t M{\"u}nchen, Theresienstra{\ss}e 37,
80333 M{\"u}nchen, Germany}
\author{Thomas Franosch}
\affiliation{Arnold Sommerfeld Center for Theoretical Physics (ASC)  and Center for
NanoScience (CeNS), Fakult{\"a}t f{\"u}r Physik,
Ludwig-Maximilians-Universit{\"a}t M{\"u}nchen, Theresienstra{\ss}e 37,
80333 M{\"u}nchen, Germany}

\begin{abstract}
Entangled networks of stiff biopolymers exhibit complex dynamic response, emerging from the topological constraints that neighboring filaments impose upon each other.
We propose a class of reference models for entanglement dynamics of stiff polymers and provide a quantitative foundation of the tube concept for stiff polymers.
For an infinitely thin needle exploring a planar course of point obstacles, we have performed large-scale computer simulations proving the conjectured scaling relations from the fast transverse equilibration to the slowest process of orientational relaxation.
We determine the rotational diffusion coefficient of the tracer, its angular confinement, the tube diameter and the orientational correlation functions.
\end{abstract}

% Statistical Physics
% 05.10.--a	computational methods
% Fluid dynamics
% 47.57.--s	complex fluids and colloidal systems
% Biological Physics
% 87.10.Tf		general theory - MD simulations


\maketitle

Semidilute solutions of stiff biopolymers form entangled networks with remarkable mechanical properties and complex dynamic response~\cite{Bausch:2006}; examples include F-actin~\cite{Liu:2006,Koenderink:2006,Wong:2004,Palmer:1999,Hinner:1998,Amblard:1996}, microtubules~\cite{Lin:2007}, the \emph{fd}~virus~\cite{Addas:2004}, and xanthan~\cite{Koenderink:2004}. Such a behavior originates from topological constraints imposed by the impenetrable neighboring filaments.
A key to the dynamics of the individual filaments is the reptation concept
pioneered by Edwards~\cite{Edwards:1967} and de Gennes~\cite{deGennes:1971}, and later extended to rods~\cite{Doi:1978}:
it summarizes the complex interaction of a single polymer with its surroundings to an effective confining tube.
Then, transport is restricted to sliding back and forth in the tube, which entails a continual remodeling of the tube ends; the resulting snake-like motion was coined ``reptation'', from the Latin \emph{repere} (to creep). Compared to the dynamics in dilute solutions, the relaxation of the polymer from its initial position and configuration becomes extremely slow.

As a consequence, the dynamic processes of entangled solutions of, e.g., biopolymers cover many decades in time,  posing a tremendous challenge both to experiments and simulations.
For flexible polymers, the reptation concept is well established~\cite{Kremer:1988+Kreer:2001} and fairly predictive~\cite{Everaers:2004+Uchida:2008};
in the case of biopolymers, only the confining tube has been observed experimentally~\cite{Kaes:1994}.
Computer simulations of entangled polymer solutions encounter major difficulties to follow the reptation motion;
yet they give insight into the relaxation within the tube~\cite{Ramanathan:2007a}.
Progress beyond simple scaling arguments depends crucially on the design of generic models, which are as simple as possible to follow the dynamics for sufficiently long times, yet complex enough to display key aspects of the underlying microscopic processes.

In this Letter, we propose the following class of models to explore single-filament transport in polymer networks: consider the motion of a tagged polymer in a plane, entangled in a course of immobilized obstacles;
the latter represent the topological constraints due to the neighboring filaments.
The reduction of dimensionality still captures the physics of entanglement since the reptation motion is essentially one-dimensional~\cite{Edwards:1967,deGennes:1971}. Stiff polymers are rather straight and therefore can be embedded in a plane, neglecting the torsion of their space curve. The orientation of the confining tube is persistent on the longest time scale of interest; thus
in video microscopy experiments, the non-trivial reptation motion of a labelled polymer found initially in the focal plane takes place in this plane.
As a benefit of the simplification, the computational complexity is lowered substantially, permitting a thorough investigation of slow dynamic processes.

This class of entanglement models sets a framework for the plethora of polymer aspects: depending on the specific problem at hand, various polymer models may be employed ranging from a flexible chain of beads and springs to a rigid rod.
It is essential to characterize and understand several limiting cases.
The physics of entanglement is singled out in the limit of hard-core interaction, vanishing width of the polymers, and zero extension of the obstacles. Then, all ramifications of excluded volume are eliminated, all configurations are permitted and equally likely, and all non-trivial dynamic correlations are due to entanglement.
In particular, the limit circumvents the  nematic phase transition.

In de Gennes's seminal paper on reptation~\cite{deGennes:1971}, a specific realization of this fundamental limit has been considered: an infinitely thin, \emph{flexible} polymer moving between fixed point obstacles. Research in the last decades rendered this model a hallmark in reptation theory. Complementary to a flexible polymer with respect to the bending stiffness is a needle, i.e., a straight, rigid object of negligible width, characterized solely by its length $L$.
In the remaining part, we focus on this important reference system.
In particular, we validate in detail the tube concept for rods~\cite{Doi:1978} and extend it towards a complete theoretical description of the rotational dynamics.

% \subsection{Needles}

\begin{figure}[b]
\centering\includegraphics[width=\linewidth]{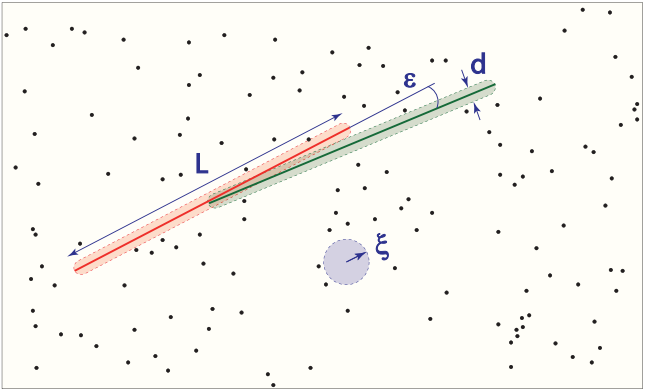}
\caption{Illustration of an entangled needle in a plane. The relevant length scales are the length of the needle $L$ and the mesh size of the network $\xi$. The surrounding point obstacles confine the needle to a tube (shaded areas) of width $d$, a renewed tube (green) is tilted against the old one (red) by an angle $\epsilon=d/L$.}
\label{fig:tube_model}
\end{figure}

For a three-dimensional suspension of needles of number density $n$, Doi and Edwards~\cite{Doi:1978} conjectured an asymptotic suppression of the rotational diffusion coefficient, $\Drot\sim n^{-2}$ as $n\to\infty$.
The slow dynamics at long times is expected to be universal irrespective of the microscopic motion; so far, research focused on ballistic needles, i.e., without solvent.
Early molecular dynamics simulations of such needle liquids~\cite{Frenkel:1981+Frenkel:1983,Magda:1986} show substantial deviations from Enskog theory. It was pointed out that the Doi-Edwards (DE) scaling of $\Drot$ is approached only very slowly and difficult to observe.
Semiquantitative agreement was found using a pseudo-dynamics~\cite{Doi:1984};
yet, the DE scaling has not been validated by a simulation of the dynamics.
Within an elaborate Boltzmann-Enskog theory~\cite{Otto:2006}, the onset of anisotropic diffusion from the dilute regime has been explained recently.
For fixed positions of the needles, the orientational degrees of freedom exhibit glassy dynamics~\cite{Schilling:2003,Renner:1995}.

% \subsection{Details of the model}

We have simulated the motion of a needle in a two-dimensional array of frozen, point-like, and hard obstacles; see~\fig{tube_model}.
The latter are distributed randomly, independently, and uniformly in the plane with an average number density~$n$. Then, the topology of the network of obstacles is characterized by the mesh size $\xi := n^{-1/2}$, i.e., the typical distance between obstacles. The degrees of freedom of the needle encompass the center-of-mass position $\vec R$ and the unit vector of orientation $\unitvec u$, the latter being parametrized by a single angle $\phi$. We compare ballistic and overdamped micro-dynamics of the needle.
In the ballistic case, the total kinetic energy is conserved, and its value sets the overall time scale $\tau_0:=L/v$ of the problem; $v$ denotes the root mean-square velocity.
For overdamped dynamics, the time scale $\tau_0$ is defined via the
coefficient of unhindered diffusion along the axis of the needle, $\tau_0=L^{2}/D_{\parallel}^0$; the free orientational diffusion is chosen in accord with hydrodynamics, $\Drot^0=6D_{\parallel}^0/L^{2}$.
In both cases, the model is specified by one single dimensionless control parameter, $n^* := n L^2$. Equivalently, the \emph{entanglement index} $\pE := -\log_{10}(\xi/L)$ quantifies the relative importance of entanglement.

% \section{Results and Discussion}

% \subsection{Rotational diffusion coefficient}

\begin{figure}
\centering\includegraphics[width=\linewidth]{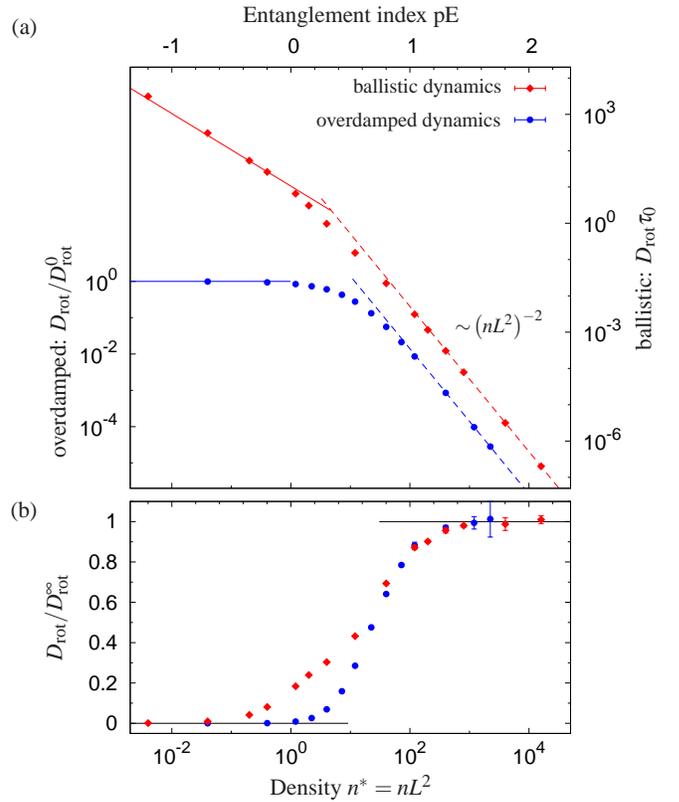}
\caption{Simulation results for the rotational diffusion coefficient.
\hspace{1em}(a)~Dashed lines are asymptotic fits to the predicted Doi-Edwards scaling, $\Drot^\infty=A (n^*)^{-2}$; the red solid line shows the result from a Boltzmann theory, $\Drot = 10.5/n^* \tau_0$~\cite{Hoefling:PhD_thesis}.
\hspace{1em}(b)~Deviation from the asymptotic behavior; note the different prefactors $A$ in $\Drot^\infty$ for ballistic and overdamped dynamics.}
\label{fig:diffusion-tube_model}
\end{figure}

\begin{figure}
\centering\includegraphics[width=\linewidth]{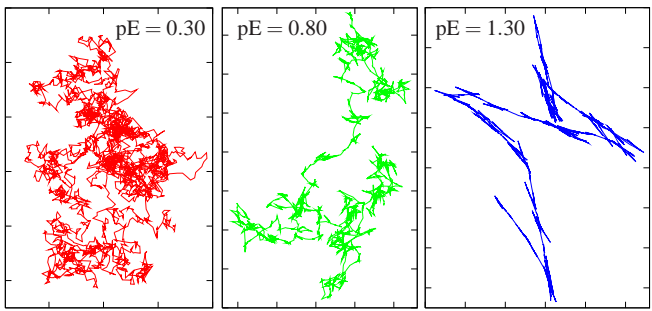}
\caption{Typical trajectories of the center of the needle for overdamped dynamics.
%; the obstacles are  omitted for clarity.
With increasing entanglement, the needle is confined to a narrowing tube, and reptation dynamics emerges; see also the Supplementary Movie~\cite{EPAPS}.}
\label{fig:traj-plots}
\end{figure}

Our simulations of the molecular dynamics are based on an event-driven algorithm using a novel approach to collision detection~\cite{Hoefling:PhD_thesis,EPAPS}.
Diffusion coefficients of the rotational motion have been extracted from the long-time behavior of the mean-square angular displacement (MSAD)  $\delta\phi^2(t):=\expect{\Delta \phi(t)^2}\simeq 2 \Drot t$ and are shown on a double-logarithmic plot in \fig{diffusion-tube_model}a.
In the investigated density range, the diffusion coefficient $\Drot$ varies over seven non-trivial decades.
For dilute systems, $n^*\ll 1$, it depends on the micro-dynamics: In the ballistic case, the diffusion coefficient is suppressed in quantitative agreement with a Boltzmann theory~\cite{Hoefling:PhD_thesis}, $\Drot = 10.5/n^* \tau_0$. For overdamped motion, the needle is unaffected by the obstacles, and the diffusion coefficient is just given by $\Drot^0$.
Once the mesh size becomes comparable to the length of the needle, $\xi\approx L$, the isotropy of rotational dynamics breaks down and a different transport mechanism develops~\cite{Moreno:2004}.
With growing entanglement, the needle is increasingly caged by the obstacle array, and eventually, its rotational motion is strongly hindered. \fig{traj-plots} illustrates the emergence of reptation-like dynamics, accompanied by a drastic suppression of the diffusion coefficient $\Drot$. For $n L^2\geq 10^2$, the data follow an asymptotic power law, $\Drot\sim (n^*)^{-2}$,  over more than four decades in the diffusion coefficient. As a most sensitive test, \fig{diffusion-tube_model}b  compares $\Drot$ to its asymptotic behavior for increasing entanglement.
Our results show that the mechanism of reptation is universal for ballistic and overdamped motion of the needle, which  will be
substantiated further in the subsequent analysis.

% \subsection{Scaling arguments}

The asymptotic suppression of $\Drot$ is rationalized by employing the concept of a confining tube~\cite{Doi:1978}. In this picture, the surrounding obstacles reduce the accessible volume of the needle to a tube of diameter~$d$ and length~$L$. The diameter is estimated from the requirement to encounter no obstacles in the tube, $d \approx 1/n L=\xi^2/L$.
The constrained motion is illustrated in \fig{traj-plots} and in a Supplementary Movie~\cite{EPAPS}: the transversal and rotational degrees of freedom are essentially frozen, permitting only displacements along the axis of the tube.
After traveling half its length, the needle is confined to a new tube tilted against the previous one on average by an angle $\epsilon\approx d/L$.
The time~$\td$ to disengage from a current tube is estimated from the free longitudinal motion inside the tube, $\td\approx \tau_0$, independent of the density.
Eventually, the orientation performs a random walk with step size $\epsilon$ and constant rate $1/\td$; hence, the diffusion coefficient scales as
\begin{equation}
\Drot \simeq \Drot^\infty := \frac{\epsilon^2}{2\td} \sim \frac{1}{n^2 L^4 \tau_0} \qquad \text{for} \quad n^* \to \infty.
\end{equation}
The given arguments apply likewise to ballistic and overdamped dynamics of the needle. In both cases,
our data in \fig{diffusion-tube_model} provide ample evidence for such a behavior, unprecedented in the literature.

\begin{figure}
\centering\includegraphics[width=\linewidth]{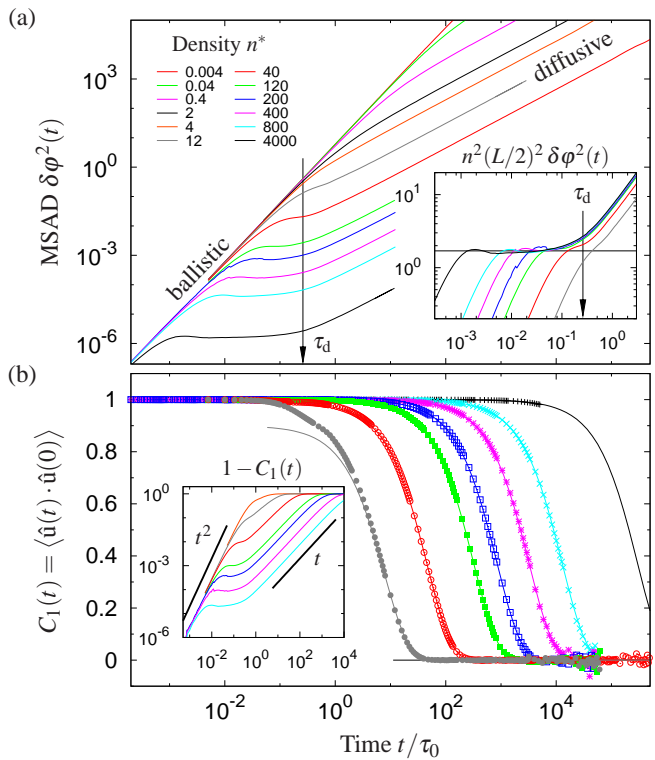}
\caption{Time dependence of the rotational motion. \hspace{1em}(a)~The mean-square angular displacement of the needle develops a plateau at high densities, which scales as $\epsilon^2\sim (n^*)^{-2}$ (cf.\ inset).
\hspace{1em}(b)~The exponential relaxation of the orientation correlation $C_1(t)$ is preceded by a plateau very close to unity (see inset). Symbols indicate simulation results and solid lines fits to $f \exp(-t/\trot)$; the fitted plateau value is consistent with the measured angular confinement, $f=1-\epsilon^2/2$.}
\label{fig:corr-msad-points}
\end{figure}

% \subsection{Mean-square angular displacement}
The quality of our data allows us to verify and quantify
the assumptions of the tube model in detail on the basis of the MSAD and correlation functions; we will restrict the discussion to ballistic dynamics. The behavior of $\delta\phi^2(t)$ is exhibited in \fig{corr-msad-points}a from the very dilute up to the highly entangled regime with $\pE>2$. At short time scales, the motion is ballistic, $\delta \phi^2(t) = \expect{\dot\phi^2} t^2$ for $t\lesssim \tcoll$, with the mean collision rate $\tau_\text{coll}^{-1}=0.845\, n^*/\tau_0$.
In the dilute regime, the MSAD directly crosses over to diffusion.
Entanglement effects emerge already at $n^*=12$ ($\pE=0.54$): when $t\approx \tcoll$, the MSAD hits an intermediate plateau, reflecting the angular confinement within the tube.
Beyond this time scale, the transverse degrees of freedom are equilibrated.
The MSAD increases further only after the tube is renewed at the time scale $\td$, and not until then, diffusion is observed.
We use the square root of the measured plateau value as definition of $\epsilon$; it follows the expected scaling law over two decades, fixing also the prefactor,  $\epsilon = 1.3 / n (L/2)^2$; see Supplementary Fig.~1 \cite{EPAPS}.
Nice data collapse is achieved for $t\gtrsim \tcoll$ by rescaling the MSAD, $(n^*)^2 \delta \phi^2(t)$; see inset of \fig{corr-msad-points}a.
The disengagement time may quantitatively be defined via extrapolation of the
diffusive asymptote to the plateau, $\epsilon^2 = 2 \Drot^\infty \td$; our data yield $\Drot^\infty =3.2 /n^2(L/2)^4\tau_0$, implying $\td = 0.3 \tau_0$ independent of the degree of entanglement.
Supplementary Fig.~1 \cite{EPAPS} also shows the tube diameter, inferred directly from an analogous plateau in the transverse mean-square displacement; we find $d=1.3\xi^2/L$ and thus $\epsilon=4.0 d/L$.

% \subsection{Orientational correlation functions}

More generally, the rotational motion in a plane is characterized in terms of orientational correlation functions
$C_\mu(t) := \expect{\!\cos [\mu\Delta \phi(t)]}$ for integer  $\mu$.
The caging inside the tube, $\delta \phi^2(t)=\epsilon^2$,
should be reflected in a plateau close to unity, $C_\mu(t) \simeq 1-\mu^2 \epsilon^2/2$ for intermediate times $\tcoll \lesssim t \lesssim \td$, which follows by a Taylor expansion of the cosine. For larger times, $t\gtrsim \td$, one expects simple
rotational diffusion; solving the diffusion equation on a circle yields $C_\mu(t) \simeq \exp(- \mu^2 \Drot t)$.
Thus, the initial orientation relaxes in two steps: fast equilibration to the plateau for $t\lesssim \tcoll$, and slow relaxation from the plateau.
We have checked that the measured $C_\mu(t)$  indeed exhibit a plateau with the predicted value, followed by
an exponential relaxation with growing decay times $\tau_\mu = 1/ \mu^2\Drot\sim (n^*)^2$.
\fig{corr-msad-points}b displays $C_1(t)$, the relaxation of the orientational persistence; this is the slowest process in the system corresponding to an equilibration time $\trot:=\tau_1=1/\Drot$.

% \section{Conclusions}

Our results unambiguously prove the conjectured scaling relations for $\Drot$, $\epsilon$, and $d$. Thus, the model reflects the generic DE~scenario, demonstrating that the essential physics due to entanglement is captured.
The predicted scaling behavior is, however, only observed in highly entangled systems with large entanglement index, $\pE \gtrsim 1$. In this regime, the trajectories indeed exhibit pronounced reptation with the typical sliding motion, see \fig{traj-plots}.

For weaker entanglement, $0< \pE \lesssim 1$, the rotational dynamics is still suppressed due to topological
constraints, but the DE~scaling is obscured by crossover phenomena.  The deviations from the predicted behavior are
highlighted in \fig{diffusion-tube_model}b by extracting the apparent amplitude of the power law.
At $\pE = 1$, where the filament length already exceeds the mesh size by a factor 10, the amplitude is still 15\% below its true asymptotic value.
One concludes that in order to observe  the scaling with an accuracy of 1\%, even stronger entanglement is required, $\pE \gtrsim 1.6$.

In real polymer solutions, the entanglement constraints are dynamically released and generated on a time scale comparable to $\td$. Within an extended model accounting for this renewal process of the obstacles, we have checked that the observed DE scaling is robust~\cite{Needle_fluctuating:2008}.

\enlargethispage{\baselineskip}

For highly entangled networks, the finite filament width in experimental situations may become relevant. Eventually, a phase transition to a nematic order occurs for long rods  at $n_\text{3d} b L^2 \simeq 1$ as has been estimated by Onsager, where $n_\text{3d}$ denotes the three-dimensional number density of rods of diameter~$b$.
The density of obstacles of our two-dimensional representation is then calculated to
$n \approx n_\text{3d} L$, hence the  nematic regime is expected for $ n^* b/L = n b L \gtrsim 1$.
For reconstituted F\hbox{-}actin solutions with filaments of $L \approx 50$\,\textmu{}m~\cite{Kaes:1994} and $b=7$\,nm,
we estimate that nematic effects are relevant only for $n^* \gtrsim 7\,000$ or $\pE\gtrsim 1.9$,
provided one can neglect the small bending flexibility. 

A finite stiffness for the polymer introduces another length scale, the persistence length, quantifying the distance over which the polymer appears as a straight rod. Due to thermal noise, there are transverse undulations which effectively blow up the width of the polymer.
It is an open question if the finite flexibility assists for the tube remodeling resulting in an enhanced rotational diffusion, or if the additional effective volume leads to further slowing down.

\begin{acknowledgments}

We are grateful to R.~Schilling for drawing our attention to the limit of point obstacles, and it is a pleasure to thank M.~Fuchs for stimulating discussions. Financial support is gratefully acknowledged
by F.H.\ from IBM Deutschland and the Nanosystems Initiative Munich,
and by T.M.\ from the Elitenetzwerk Bayern.

\end{acknowledgments}

% \bibliographystyle{apsrev}
% \bibliography{needle}

\begin{thebibliography}{30}
\providecommand{\natexlab}[1]{#1}
\providecommand{\bibnamefont}[1]{#1}
\providecommand{\bibfnamefont}[1]{#1}
\providecommand{\citenamefont}[1]{#1}
\providecommand{\bibinfo}[2]{#2}

\bibitem[{\citenamefont{Bausch and Kroy}(2006)}]{Bausch:2006}
\bibinfo{author}{\bibfnamefont{A.~R.} \bibnamefont{Bausch}} \bibnamefont{and}
  \bibinfo{author}{\bibfnamefont{K.}~\bibnamefont{Kroy}},
  \bibinfo{journal}{Nature Phys.} \textbf{\bibinfo{volume}{2}},
  \bibinfo{pages}{231} (\bibinfo{year}{2006}).

\bibitem[{\citenamefont{Liu} \emph{et~al.}(2006)}]{Liu:2006}
\bibinfo{author}{\bibfnamefont{J.}~\bibnamefont{Liu}} \emph{et~al.},
  \bibinfo{journal}{Phys. Rev. Lett.} \textbf{\bibinfo{volume}{96}},
  \bibinfo{pages}{118104} (\bibinfo{year}{2006}).

\bibitem[{\citenamefont{Koenderink} \emph{et~al.}(2006)}]{Koenderink:2006}
\bibinfo{author}{\bibfnamefont{G.~H.} \bibnamefont{Koenderink}} \emph{et~al.},
  \bibinfo{journal}{Phys. Rev. Lett.} \textbf{\bibinfo{volume}{96}},
  \bibinfo{eid}{138307} (\bibinfo{year}{2006}).

\bibitem[{\citenamefont{Wong} \emph{et~al.}(2004)}]{Wong:2004}
\bibinfo{author}{\bibfnamefont{I.~Y.} \bibnamefont{Wong}} \emph{et~al.},
  \bibinfo{journal}{Phys. Rev. Lett.} \textbf{\bibinfo{volume}{92}},
  \bibinfo{eid}{178101} (\bibinfo{year}{2004}).

\bibitem[{\citenamefont{Palmer} \emph{et~al.}(1999)}]{Palmer:1999}
\bibinfo{author}{\bibfnamefont{A.}~\bibnamefont{Palmer}} \emph{et~al.},
  \bibinfo{journal}{Biophys. J.} \textbf{\bibinfo{volume}{76}},
  \bibinfo{pages}{1063} (\bibinfo{year}{1999}).

\bibitem[{\citenamefont{Hinner} \emph{et~al.}(1998)}]{Hinner:1998}
\bibinfo{author}{\bibfnamefont{B.}~\bibnamefont{Hinner}} \emph{et~al.},
  \bibinfo{journal}{Phys. Rev. Lett.} \textbf{\bibinfo{volume}{81}},
  \bibinfo{pages}{2614} (\bibinfo{year}{1998}).

\bibitem[{\citenamefont{Amblard} \emph{et~al.}(1996)}]{Amblard:1996}
\bibinfo{author}{\bibfnamefont{F.}~\bibnamefont{Amblard}} \emph{et~al.},
  \bibinfo{journal}{Phys. Rev. Lett.} \textbf{\bibinfo{volume}{77}},
  \bibinfo{pages}{4470} (\bibinfo{year}{1996}).

\bibitem[{\citenamefont{Lin} \emph{et~al.}(2007)}]{Lin:2007}
\bibinfo{author}{\bibfnamefont{Y.}~\bibnamefont{Lin}} \emph{et~al.},
  \bibinfo{journal}{Macromolecules} \textbf{\bibinfo{volume}{40}},
  \bibinfo{pages}{7714} (\bibinfo{year}{2007}).

\bibitem[{\citenamefont{Addas} \emph{et~al.}(2004)\citenamefont{Addas, Schmidt,
  and Tang}}]{Addas:2004}
\bibinfo{author}{\bibfnamefont{K.~M.} \bibnamefont{Addas}},
  \bibinfo{author}{\bibfnamefont{C.~F.} \bibnamefont{Schmidt}},
  \bibnamefont{and} \bibinfo{author}{\bibfnamefont{J.~X.} \bibnamefont{Tang}},
  \bibinfo{journal}{Phys. Rev. E} \textbf{\bibinfo{volume}{70}},
  \bibinfo{eid}{021503} (\bibinfo{year}{2004}).

\bibitem[{\citenamefont{Koenderink} \emph{et~al.}(2004)}]{Koenderink:2004}
\bibinfo{author}{\bibfnamefont{G.~H.} \bibnamefont{Koenderink}} \emph{et~al.},
  \bibinfo{journal}{Phys. Rev. E} \textbf{\bibinfo{volume}{69}},
  \bibinfo{eid}{021804} (\bibinfo{year}{2004}).

\bibitem[{\citenamefont{Edwards}(1967)}]{Edwards:1967}
\bibinfo{author}{\bibfnamefont{S.~F.} \bibnamefont{Edwards}},
  \bibinfo{journal}{Proc. Phys. Soc.} \textbf{\bibinfo{volume}{92}},
  \bibinfo{pages}{9} (\bibinfo{year}{1967}).

\bibitem[{\citenamefont{de~Gennes}(1971)}]{deGennes:1971}
\bibinfo{author}{\bibfnamefont{P.~G.} \bibnamefont{de~Gennes}},
  \bibinfo{journal}{J.~Chem. Phys.} \textbf{\bibinfo{volume}{55}},
  \bibinfo{pages}{572} (\bibinfo{year}{1971}).

\bibitem[{\citenamefont{Doi and Edwards}(1978)}]{Doi:1978}
\bibinfo{author}{\bibfnamefont{M.}~\bibnamefont{Doi}} \bibnamefont{and}
  \bibinfo{author}{\bibfnamefont{S.~F.} \bibnamefont{Edwards}},
  \bibinfo{journal}{J.~Chem. Soc., Faraday Trans.~2}
  \textbf{\bibinfo{volume}{74}}, \bibinfo{pages}{560} (\bibinfo{year}{1978}).

\bibitem[{\citenamefont{Kremer} \emph{et~al.}(1988)\citenamefont{Kremer, Grest,
  and Carmesin}}]{Kremer:1988+Kreer:2001}
\bibinfo{author}{\bibfnamefont{K.}~\bibnamefont{Kremer}},
  \bibinfo{author}{\bibfnamefont{G.~S.} \bibnamefont{Grest}}, \bibnamefont{and}
  \bibinfo{author}{\bibfnamefont{I.}~\bibnamefont{Carmesin}},
  \bibinfo{journal}{Phys. Rev. Lett.} \textbf{\bibinfo{volume}{61}},
  \bibinfo{pages}{566} (\bibinfo{year}{1988});
% \bibitem[{\citenamefont{Kreer} \emph{et~al.}(2001)}]{Kreer:2001}
\bibinfo{author}{\bibfnamefont{T.}~\bibnamefont{Kreer}} \emph{et~al.},
  \bibinfo{journal}{Macromolecules} \textbf{\bibinfo{volume}{34}},
  \bibinfo{pages}{1105} (\bibinfo{year}{2001}).

\bibitem[{\citenamefont{Everaers} \emph{et~al.}(2004)}]{Everaers:2004+Uchida:2008}
\bibinfo{author}{\bibfnamefont{R.}~\bibnamefont{Everaers}} \emph{et~al.},
  \bibinfo{journal}{Science} \textbf{\bibinfo{volume}{303}},
  \bibinfo{pages}{823} (\bibinfo{year}{2004});
% \bibitem[{\citenamefont{Uchida} \emph{et~al.}(2008)\citenamefont{Uchida, Grest,
%   and Everaers}}]{Uchida:2008}
\bibinfo{author}{\bibfnamefont{N.}~\bibnamefont{Uchida}},
  \bibinfo{author}{\bibfnamefont{G.~S.} \bibnamefont{Grest}}, \bibnamefont{and}
  \bibinfo{author}{\bibfnamefont{R.}~\bibnamefont{Everaers}},
  \bibinfo{journal}{J. Chem. Phys.} \textbf{\bibinfo{volume}{128}},
  \bibinfo{eid}{044902} (\bibinfo{year}{2008}).

\bibitem[{\citenamefont{K{\"a}s} \emph{et~al.}(1994)\citenamefont{K{\"a}s,
  Strey, and Sackmann}}]{Kaes:1994}
\bibinfo{author}{\bibfnamefont{J.}~\bibnamefont{K{\"a}s}},
  \bibinfo{author}{\bibfnamefont{H.}~\bibnamefont{Strey}}, \bibnamefont{and}
  \bibinfo{author}{\bibfnamefont{E.}~\bibnamefont{Sackmann}},
  \bibinfo{journal}{Nature} \textbf{\bibinfo{volume}{368}},
  \bibinfo{pages}{226} (\bibinfo{year}{1994}).

\bibitem[{\citenamefont{Ramanathan and Morse}(2007)}]{Ramanathan:2007a}
\bibinfo{author}{\bibfnamefont{S.}~\bibnamefont{Ramanathan}} \bibnamefont{and}
  \bibinfo{author}{\bibfnamefont{D.~C.} \bibnamefont{Morse}},
  \bibinfo{journal}{Phys. Rev. E} \textbf{\bibinfo{volume}{76}},
  \bibinfo{eid}{010501} (\bibinfo{year}{2007}).

\bibitem[{\citenamefont{Frenkel and Maguire}(1981)}]{Frenkel:1981+Frenkel:1983}
\bibinfo{author}{\bibfnamefont{D.}~\bibnamefont{Frenkel}} \bibnamefont{and}
  \bibinfo{author}{\bibfnamefont{J.~F.} \bibnamefont{Maguire}},
  \bibinfo{journal}{Phys. Rev. Lett.} \textbf{\bibinfo{volume}{47}},
  \bibinfo{pages}{1025} (\bibinfo{year}{1981});
% \bibitem[{\citenamefont{Frenkel and Maguire}(1983)}]{Frenkel:1983}
% \bibinfo{author}{\bibfnamefont{D.}~\bibnamefont{Frenkel}} \bibnamefont{and}
%   \bibinfo{author}{\bibfnamefont{J.~F.} \bibnamefont{Maguire}},
  \bibinfo{journal}{Mol. Phys.} \textbf{\bibinfo{volume}{49}},
  \bibinfo{pages}{503} (\bibinfo{year}{1983}).

\bibitem[{\citenamefont{Magda} \emph{et~al.}(1986)\citenamefont{Magda, Davis,
  and Tirrell}}]{Magda:1986}
\bibinfo{author}{\bibfnamefont{J.~J.} \bibnamefont{Magda}},
  \bibinfo{author}{\bibfnamefont{H.~T.} \bibnamefont{Davis}}, \bibnamefont{and}
  \bibinfo{author}{\bibfnamefont{M.}~\bibnamefont{Tirrell}},
  \bibinfo{journal}{J.~Chem. Phys.} \textbf{\bibinfo{volume}{85}},
  \bibinfo{pages}{6674} (\bibinfo{year}{1986}).

\bibitem[{\citenamefont{Doi} \emph{et~al.}(1984)\citenamefont{Doi, Yamamoto,
  and Kano}}]{Doi:1984}
\bibinfo{author}{\bibfnamefont{M.}~\bibnamefont{Doi}},
  \bibinfo{author}{\bibfnamefont{I.}~\bibnamefont{Yamamoto}}, \bibnamefont{and}
  \bibinfo{author}{\bibfnamefont{F.}~\bibnamefont{Kano}},
  \bibinfo{journal}{J.~Phys. Soc. Jpn.} \textbf{\bibinfo{volume}{53}},
  \bibinfo{pages}{3000} (\bibinfo{year}{1984}).

\bibitem[{\citenamefont{Otto} \emph{et~al.}(2006)\citenamefont{Otto,
  Aspelmeier, and Zippelius}}]{Otto:2006}
\bibinfo{author}{\bibfnamefont{M.}~\bibnamefont{Otto}},
  \bibinfo{author}{\bibfnamefont{T.}~\bibnamefont{Aspelmeier}},
  \bibnamefont{and}
  \bibinfo{author}{\bibfnamefont{A.}~\bibnamefont{Zippelius}},
  \bibinfo{journal}{J. Chem. Phys.} \textbf{\bibinfo{volume}{124}},
  \bibinfo{eid}{154907} (\bibinfo{year}{2006}).

\bibitem[{\citenamefont{Schilling and Szamel}(2003)}]{Schilling:2003}
\bibinfo{author}{\bibfnamefont{R.}~\bibnamefont{Schilling}} \bibnamefont{and}
  \bibinfo{author}{\bibfnamefont{G.}~\bibnamefont{Szamel}},
  \bibinfo{journal}{Europhys. Lett.} \textbf{\bibinfo{volume}{61}},
  \bibinfo{pages}{207} (\bibinfo{year}{2003}).

\bibitem[{\citenamefont{Renner} \emph{et~al.}(1995)\citenamefont{Renner,
  L\"owen, and Barrat}}]{Renner:1995}
\bibinfo{author}{\bibfnamefont{C.}~\bibnamefont{Renner}},
  \bibinfo{author}{\bibfnamefont{H.}~\bibnamefont{L\"owen}}, \bibnamefont{and}
  \bibinfo{author}{\bibfnamefont{J.~L.} \bibnamefont{Barrat}},
  \bibinfo{journal}{Phys. Rev. E} \textbf{\bibinfo{volume}{52}},
  \bibinfo{pages}{5091} (\bibinfo{year}{1995}).

\bibitem[{\citenamefont{H{\"o}f\/ling}(2006)}]{Hoefling:PhD_thesis}
\bibinfo{author}{\bibfnamefont{F.}~\bibnamefont{H{\"o}f\/ling}}, Ph.D. thesis,
  \bibinfo{school}{Ludwig-Maximilians-Universit{\"a}t M{\"u}nchen}
  (\bibinfo{year}{2006}), \bibinfo{note}{{ISBN:} 978-3-86582-426-4}.

\bibitem[{EPA()}]{EPAPS}
  \bibinfo{note}{{S}ee EPAPS Document No. [to be inserted]. For more
  information on EPAPS, see http://www.aip.org/pubservs/epaps.html.}

\bibitem[{\citenamefont{Moreno and Kob}(2004)}]{Moreno:2004}
\bibinfo{author}{\bibfnamefont{A.~J.} \bibnamefont{Moreno}} \bibnamefont{and}
  \bibinfo{author}{\bibfnamefont{W.}~\bibnamefont{Kob}},
  \bibinfo{journal}{Europhys. Lett.} \textbf{\bibinfo{volume}{67}},
  \bibinfo{pages}{820} (\bibinfo{year}{2004}).

\bibitem[{\citenamefont{H{\"o}f\/ling}
  \emph{et~al.}()\citenamefont{H{\"o}f\/ling, Munk, Frey, and
  Franosch}}]{Needle_fluctuating:2008}
\bibinfo{author}{\bibfnamefont{F.}~\bibnamefont{H{\"o}f\/ling}},
  \bibinfo{author}{\bibfnamefont{T.}~\bibnamefont{Munk}},
  \bibinfo{author}{\bibfnamefont{E.}~\bibnamefont{Frey}}, \bibnamefont{and}
  \bibinfo{author}{\bibfnamefont{T.}~\bibnamefont{Franosch}},
  \bibinfo{note}{in preparation}.

\end{thebibliography}

\end{document}